\documentclass[twocolumn,preprintnumbers,superscriptaddress,nofootinbib,aps,prd,floatfix]{revtex4}

\usepackage{footmisc,multirow}
\usepackage{subfigure}
\usepackage{amsmath,slashed,enumerate,color}

\usepackage{graphicx}

\newcommand{\be}{\begin{eqnarray*}}
\newcommand{\ee}{\end{eqnarray*}}

\newcommand{\bee}{\begin{eqnarray}}
\newcommand{\eee}{\end{eqnarray}}
\newcommand{\beeq}{\begin{equation}}
\newcommand{\eeeq}{\end{equation}}

\newcommand{\BR}{\text{BR}}

\newcommand{\sh}{\hat{s}}
\newcommand{\abs}[1]{\left| #1 \right|}

\preprint{IPPP/14/91} 
\preprint{DCPT/14/182}

\begin{document}

\title{Off-Shell Higgs Coupling Measurements in BSM scenarios}

\begin{abstract}
  Proposals of measuring the off-shell Higgs contributions and first
  measurements at the LHC have electrified the Higgs phenomenology
  community for two reasons: Firstly, probing interactions at high
  invariant masses and momentum transfers is intrinsically sensitive
  to new physics beyond the Standard Model, irrespective of a resonant
  or non-resonant character of a particular BSM scenario. Secondly,
  under specific assumptions a class of models exists for which the
  off-shell coupling measurement together with a measurement of the
  on-shell signal strength can be re-interpreted in terms of a bound
  on the total Higgs boson width.  In this paper, we provide a first
  step towards a classification of the models for which a total width
  measurement is viable and we discuss examples of BSM models for
  which the off-shell coupling measurement can be important in either
  constraining or even discovering new physics in the upcoming LHC
  runs. Specifically, we discuss the quantitative impact of the
  presence of dimension six operators on the (de)correlation of Higgs
  on- and off-shell regions keeping track of all interference
  effects. We furthermore investigate off-shell measurements in a
  wider context of new (non-)resonant physics in Higgs portal
  scenarios and the MSSM.
\end{abstract}

\author{Christoph Englert} \email{christoph.englert@glasgow.ac.uk}
\affiliation{SUPA, School of Physics and Astronomy,\\University of
  Glasgow, Glasgow G12 8QQ, United Kingdom\\[0.1cm]}

\author{Yotam Soreq} \email{yotam.soreq@weizmann.ac.il}
\affiliation{Department of Particle Physics and
  Astrophysics,\\Weizmann Institute of Science, Rehovot 7610001,
  Israel\\[0.1cm]}

\author{Michael Spannowsky}\email{michael.spannowsky@durham.ac.uk} 
\affiliation{Institute for
  Particle Physics Phenomenology, Department of Physics,\\Durham
  University, Durham DH1 3LE, United Kingdom\\[0.1cm]}

\maketitle

\section{Introduction}
\label{sec:intro}
The Higgs discovery in 2012~\cite{hatlas,hcms} with subsequent (rather
inclusive) measurements performed in agreement with the Standard Model
(SM) hypothesis~\cite{theo,hzz} highlight the necessity to establish
new Higgs physics-related search and analysis strategies that are
sensitive to beyond the SM (BSM) interactions. In a phenomenological
bottom-up approach the LHC's sensitivity reach can be used to classify
potential BSM physics, which we can loosely categorize models into
four classes:
\begin{equation}
  \label{eq:states}
  \hspace{-0.1cm}
  \parbox{0.43\textwidth}{\vspace{-0.4cm}
    \begin{enumerate}[(i)]
    \item light hidden degrees of freedom,
    \item new degrees of freedom in the sub-TeV that induce non-resonant
      thresholds,
    \item resonant TeV scale degrees of freedom with parametrically suppressed
      production cross sections,
    \item new degrees of freedom in the multi-TeV range that can be probed
      in the energetic tail region of the 13 and 14 TeV options, or might
      even lie outside the energetic coverage of the LHC.
    \end{enumerate}
    \vspace{-0.3cm}
  }
\end{equation}
The analysis strategies with which the LHC multi-purpose experiments
can look for an individual category above typically build upon
assumptions about the remaining three. These assumptions need to be
specified in order for the result to have potential interpretation
beyond the limitations of a certain specified scenario.

\bigskip

For example, if we deal with a large hierarchy of physics scales as in
case~(iv), we can rely on effective theory methods to set limits on
the presence of new scale-separated dynamics. A well-motivated
approach in light of electroweak precision measurements and current
Higgs analyses is to extend the renormalizable SM Lagrangian by
dimension six
operators~\cite{dim6,dim6e,dim6r,dim6g,dim6gg1,dim6gg2,dim6f}, which
parametrize the leading order corrections of SM dynamics in the
presence of new heavy states model-independently.

Given that the LHC machine marginalizes over a vast partonic energy
range, the described effective field theory~(EFT) methods are not
applicable in cases~(i)-(iii), for which new resonant dynamics is
resolved; we cannot trust an EFT formulation in the presence of
thresholds. In these cases we have to rely on agreed benchmark
scenarios to make the interpretation of a limit setting exercise
transparent.

\begin{figure*}[!t]
  \includegraphics[width=0.75\textwidth]{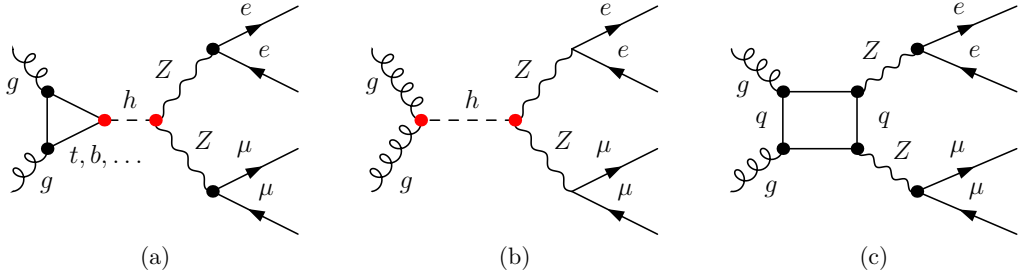}
  \caption{\label{fig:feyn} Representative Feynman diagram topologies
    contributing to $gg\to ZZ\to e^+ e^- \mu^+\mu^-$. Additional
    particles can run in the Higgs production loops (a)
    (Sec.~\ref{sec:ldof}), (b) the Higgs vertices can be modified by
    higher dimensional operator contributions (Sec.~\ref{sec:eft}), or
    additional $s$-channel resonances can show up with $m_\phi>m_h$
    (Sec.~\ref{sec:res}).}
\end{figure*}

In general, the standard analysis approach to BSM scenarios that fall
into categories~(ii)-(iv) focuses on large invariant masses and large
momentum transfers. However, it is intriguing that a correlation of
the low and high invariant mass measurements also allows us to
constrain scenarios of type~(i). An important analysis that has
received a lot of attention from both the theoretical and the
experimental community in this regard is the Higgs width measurement
in $pp\to ZZ\to 4\ell$ as introduced by Caola and
Melnikov~\cite{melnikov}. Assuming the SM spectrum and neglecting
renormalizability issues that arise when we employ the
$\kappa$-language of recent Higgs coupling
measurements~\cite{HiggsXSWG}, the proposed strategy exploits
non-decoupling of the top loop contributing to $pp\to h\to ZZ$
(directly related to the top mass' generation via the Higgs mechanism)
and decoupling of the Higgs width parameter for large invariant $ZZ$
masses to formulate a constraint on the Higgs width:
\begin{subequations}
  \begin{alignat}{4}
    \label{eq:1a}
    \mu^{\text{on}}_{ZZ} \equiv& \frac{\sigma_{h}\times \BR(h\to ZZ\to
      4\ell)}{\left[ \sigma_{h}\times \BR(h\to ZZ\to 4\ell)
      \right]_{\rm SM}} \sim& {\kappa^2_{ggh}\, \kappa^2_{hZZ}
      \over \Gamma_h/\Gamma_h^{\rm SM}}\,,\\
    \label{eq:1b}
    \mu^{\text{off}}_{ZZ}\equiv& \frac{ {\text{d}}\overline \sigma_h
    }{ \left[ {\text{d}}\overline \sigma_h \right]_{\rm SM} } \sim
    \kappa^2_{ggh}(\sh) \, \kappa^2_{hZZ} (\sh) \, ,
  \end{alignat}
\end{subequations}
where $\sqrt{\sh}$ is the partonic level center of mass energy and
$\kappa_X\equiv (g_X + \tilde g _X) /g_X$, where $g_X$ is the coupling
in the SM and $\tilde g $ parametrizes BSM effects. Here, For
simplicity, here we only consider gluon fusion, the dominant
production mechanism. ``Off-shell'' typically means $m_{ZZ}\gtrsim
330$~GeV due to a maximized ratio of Higgs-induced vs. continuum
$gg\to ZZ$ production as a consequence of the top threshold.

If we have $\Gamma_h>\Gamma_h^{\text{SM}}\simeq 4~\text{MeV}$, yet
still a SM value for the $pp\to h\to ZZ$ signal strength
$\mu^{\text{on}}_{ZZ}$, we need to have $\kappa^2_{ggh}\,
\kappa^2_{hZZ}>1$. If we consider an extrapolation of the on-shell
region to the off-shell region based on the SM Feynman graph templates
depicted in Fig.~\ref{fig:feyn}, we can understand a constraint on
$\overline\sigma_h $ as a constraint on $\Gamma_h$ {\emph{as a
    consistency check}}: In a well-defined QFT framework such as the
SM, a particle width is a consequence of the interactions and degrees
of freedom as specified in the Lagrangian density. E.g. by extending
the SM with dynamics that induce an invisible partial Higgs decay
width, there is no additional information in the off-shell measurement
when combined with the on-shell signal strength. It is important to
note that if we observe an excess in $\overline\sigma_h$ in the
future, then this will not be a manifestation of
$\Gamma_h>\Gamma_h^{\text{SM}}$. Instead we will necessarily have to
understand this as a observation of physics beyond the SM, which might
but does not need to be in relation to the Higgs boson.

A quantitatively correct estimate of important interference effects
that shape $\overline \sigma_h$ have been provided in
Refs.~\cite{Kauer:2013qba,ciaran,mit} (see also~\cite{Dixon:2013haa2}
for a related discussion of $pp\to h\to \gamma\gamma$). These
interference effects are an immediate consequence of a well-behaved
electroweak sector in the sub-TeV range in terms of renormalizability
and, hence, unitarity~\cite{bsm,Chanowitz:1978mv}. While they remain
calculable in electroweak leading order Monte Carlo
programs~\cite{Kauer:2013qba,ciaran}, they are not theoretically
well-defined, unless we assume a specific BSM scenario or invoke EFT
methods. For a discussion on the unitarity constraints on the
different Wilson coefficients see~\cite{kramer}.

Both ATLAS and CMS have performed the outlined measurement with the 8
TeV data set in the meantime~\cite{cmswidth,atlaswidth}. The
importance of high invariant mass measurements in this particular
channel in a wider context has been discussed
in~Refs.~\cite{fermilab1,fermilab2,bsm,bsmh,bsmh2}

In the particular case of $pp\to ZZ\to 4\ell$, we can classify models
according to their effect in the on-shell and off-shell phase space
regions.  We can identify four regions depending on the measured value
of $\mu^{\text{off}}_{ZZ}$, which can provide a strong hint for new
physics in the above scenarios (ii)-(iv):
\begin{equation}
  \label{eq:interp}
  \parbox{6cm}{\vspace{-0.4cm}
  \begin{enumerate}
  \item $\mu^{\text{off}}_{ZZ} = 1$ and $[\kappa^2_{ggh}\kappa^2_{hZZ}]^{\text{on}} = 1$\,,
  \item $\mu^{\text{off}}_{ZZ} = 1$ and $[\kappa^2_{ggh}\kappa^2_{hZZ}]^{\text{on}} \neq 1$\,,
  \item $\mu^{\text{off}}_{ZZ} \neq 1$ and $[\kappa^2_{ggh}\kappa^2_{hZZ}]^{\text{on}} = 1$\,,
  \item $\mu^{\text{off}}_{ZZ} \neq 1$ and $[\kappa^2_{ggh}\kappa^2_{hZZ}]^{\text{on}} \neq 1$\,.
  \end{enumerate} 
  \vspace{-0.4cm}
}
\end{equation}

We can write a generalized version of Eq.~(\ref{eq:1b}) that also
reflects (non-)resonant BSM effects by writing the general
amplitude
\begin{multline}
  \label{eq:ggzz}
  {\cal{M}}(gg\to ZZ) = \bigg[ [g_{hZZ} g_{ggh}](\hat s, \hat t) + [\tilde
  g_{hZZ} \tilde g_{ggh} ](\hat s, \hat t) \\
  +\sum_i [\tilde g_{ggX_i}
  \tilde g_{X_iZZ}](\hat s, \hat t)\bigg]\\
  + \big\{ g_{ggZZ}(\hat s, \hat t) + \tilde g_{ggZZ} (\hat s, \hat t) \big\}\,,
\end{multline}
from which we may compute $\hbox{d} \overline \sigma(gg)\sim
|{\cal{M}}|^2$ by folding with parton distribution functions and the
phase space weight. For $\bar q q$-induced $ZZ$ production we can
formulate a similar amplitude
\begin{multline}
  {\cal{M}}(\bar qq\to ZZ) = g_{\bar q q ZZ} (\hat s, \hat t) \\ +
  \tilde g_{\bar q q ZZ} (\hat s, \hat t) + \sum_i \tilde [g_{\bar q q
    X_i} \tilde g_{X_iZZ}](\hat s, \hat t)\,,
\end{multline}
which can impact the $Z$ boson pair phenomenology on top of the
$gg$-induced channels. Hence, for the differential off-shell cross
section we find $\hbox{d} \overline \sigma \simeq \hbox{d} \overline
\sigma(gg) + \hbox{d} \overline \sigma(\bar q q)$.

Resonant scenarios, such as new scalars and vectors are in agreement with
the generalized Landau-Yang theorem~\cite{landau} have been studied in
detail~\cite{more}.  Non-resonant new interactions involving light
quarks, e.g. in a dimension six operator extension of the SM, are
typically constrained.

For all models that fall into the classification 1. we are allowed to
re-interpret the off-shell measurement as a constraint on the Higgs
width bearing in mind theoretical shortcomings when parameters are
varied inconsistently; the uncertainty of a measurement of
$\mu^{\text{off}}_{ZZ}$ and the on-shell signal strength
$\mu^{\text{on}}_{ZZ}$ combine to a constraint on $\Gamma_h$. Assuming
new physics exists, such a constraint makes strong assumptions about
potential cancellations among or absence of the new physics couplings
in the off-shell region. In particular because the effective couplings
are phase space dependent and can affect the differential $m_{ZZ}$
distribution beyond a simple rescaling.  A concrete example of this
class of models is the general dimension six extension of the SM Higgs
sector with a Higgs portal to provide an invisible partial decay width
$\Gamma^{\text{inv}}$. If we are in the limit of vanishing dimension
six Wilson coefficients $c_i\ll v^2/f^2$, new EFT physics
contributions with new physics scale $f$ in the on- and off-shell
regions are parametrically suppressed and the dominant unconstrained
direction in this measurement is $\Gamma^{\text{inv}}$. Note that
there can be cancellations in the high invariant mass region among
different dimension six coefficients, so the constraint formulated on
$\Gamma^{\text{inv}}$ \emph{requires} $c_i\to 0$.

For the second scenario a re-interpretation in terms of a width
measurement is generally not valid. Here, the SM off-shell
distribution is recovered while the on-shell signal strength is unity
due to a cancellation between the modified Higgs width and the
on-shell coupling modification. A toy-model example has been discussed
in~\cite{bsm}.

From a phenomenological point of view, scenarios 3. and 4. are of
great interest, in particular because SM-like signal strength
measurements alone do typically not provide enough information to rule
out models conclusively. Most concrete realizations of BSM physics
predict new physics at high energies as a unitarity-related
compensator for modifications of on-shell coupling
strengths. ``Off-shell'' measurements are therefore prime candidates
to look for deviations from the Standard Model in the sense that they
will be sensitive to new resonances~\cite{bsmres} and will have strong
implications for BSM physics in general.

The aim of this work is to provide a survey of the reach of the
validity of the Higgs width interpretation. Since modifications of the
Higgs width do imply physics beyond the SM, the Higgs width
interpretation can be reconciled with new physics effects in the $ZZ$
channel. This allows us to make contact to concrete phenomenological
realizations using the above categorization. New degrees of freedom as
introduced in the beginning of this section that give rise to new
contributions following Eq.~\eqref{eq:ggzz}.

We focus on $gg$-induced $ZZ$ production throughout. We will first
discuss light non-resonant degrees of freedom and their potential
impact on the $m_{ZZ}$ distributions with the help of toy models that
we generalize to the (N)MSSM in Sec.~\ref{sec:ldof}. Assuming a scale
separation between new resonant phenomena and the probed energy scales
in $pp\to ZZ \to 4\ell$ we discuss high invariant mass $Z$ boson pair
production in a general dimension six extension of the SM in
Sec.~\ref{sec:eft} before we consider resonant phenomena in
Sec.~\ref{sec:res}.  In particular, our calculation includes all
interference effects (at leading order) of $pp\to ZZ\to 4\ell$ in all
of these scenarios. Our discussions and findings straightforwardly
apply to the $WW$ channel which is, due to custodial symmetry, closely
related to the $ZZ$ final state.

\section{A note on the Monte Carlo Implementation}

The numerical calculations in this paper have been obtained with a
customized version of {\sc{Vbfnlo}}~\cite{vbfnlo}, that
employs~{\sc{FeynArts/FormCalc/LoopTools}}~\cite{feyntools} tool chain
for the full $pp\to ZZ \to e^+e^-\mu^+\mu^-$ final state (see
Fig.~\ref{fig:feyn}). We neglect QED contributions throughout; they
are known to be negligible especially for the high $m_{ZZ}$ phase
space region where both $Z$ bosons can be fully reconstructed. Our
implementation is detailed in~\cite{bsm} and has been validated
against the SM results of~\cite{ciaran}. We include bottom quark
contributions to the Higgs diagrams in Fig.~\ref{fig:feyn}, these can
become relevant in the MSSM at large $\tan\beta$. The effective theory
implementation has been checked for consistency against existing
implementations~\cite{Alloul:2013naa} (normalizations and Feynman
rules) based on {\sc{FeynRules}}~\cite{feynrules}. The phase space
integration has been validated against the results of
\cite{ciaran}. Throughout we apply inclusive cuts
\begin{multline}
  \Delta R_{\ell\ell'} \geq 0.4,~|y_\ell | \leq 2.5,~p_{{\rm T},\ell}\geq
  10~\text{GeV}\, ,
\end{multline}
where $ \Delta R_{\ell\ell'}$ is the angular separation between any
two leptons, $y_\ell$ and $p_{{\rm T},\ell}$ are the lepton rapidity
and transverse momentum respectively, and focus on LHC collisions at
13~TeV.

\section{Non-Resonant BSM Physics}
\label{sec:nonres}

\subsection*{Qualitative discussion of BSM contributions}
\label{sec:qualnonres}

To zoom in on the classes of models where a width interpretation is
valid we note that, assuming peculiar cancellation effects among the
couplings are absent, the coupling which has to be present and affects
the on-shell and off-shell region in the least constraint way is the
$ggh$ coupling. Further, crucial to a width interpretation
in~\eqref{eq:interp} is a strict correlation of the on- and off-shell
regions which can be broken if light degrees of freedom are present
following our classification in~\eqref{eq:states}. If these light
states carry color charge and obtain a mass that is unrelated to the
electroweak vacuum, they will decouple quickly for $m_{ZZ}\gg m_h$,
although they can provide a notable contribution to the Higgs on-shell
region~\cite{bsm}. Inspired by the assumption that $\kappa_i^{\rm
  on}=\kappa_i^{\rm off}$~\cite{atlaswidth}, parametrically this
correlation requirement for $ggh$ is captured by the complex double
ratio
\begin{equation}
  \label{eq:doubler}
  R(m_{ZZ})=\kappa_{ggh}(m_{ZZ}^2)/\kappa_{ggh}(m_h^2)\,.
\end{equation}
If $R\simeq 1$ \emph{independent of $m_{ZZ}$} within experimental
uncertainties, the off-shell coupling measurement can be
re-interpreted in terms of a width measurement. Note,
$\mu^{\mathrm{on}}_{ZZ}=1$ has to be imposed as an additional
requirement to ensure consistency with experimental
measurements. Scenarios 1. and 4. can satisfy this condition, however,
if a significant deviation of the Standard Model prediction is
observed in the off-shell regime reinterpreting this observation in
terms of a non-SM-like width for the Higgs resonance is likely to be
of minor interest compared to the discovery of new physics.

The (de)correlation between the on- and off-shell measurements can be
demonstrated by the following simple toy examples: we consider a
scalar $S$ with mass $m_s$, a fermion $f$ with mass $m_f$ as extra
particles added to the SM spectrum. We allow these states to couple to
the Higgs boson with interactions
\begin{align}
\label{eq:toy}
  {\cal L}_{\rm toy} = -c_s \frac{2m^2_s}{v} hS^\dagger S - c_f
  \frac{m_f}{v}h \bar{f} f \, ,
\end{align}
where $v\simeq 246$~GeV. The coefficients $c_{f,s}$ parameterize the
deviation from the SM-like case where the entire particle mass is
originated from the Higgs mechanism with one doublet. In addition, we
also take into account the contribution of the dimension six operator
$H^\dagger H G_{\mu\nu}^a G^{a\mu\nu}\,$.

\begin{figure}[!t]
  \begin{center}
    \includegraphics[width=0.45\textwidth]{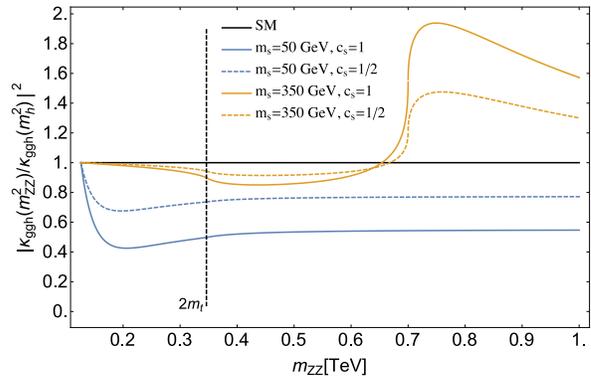}
    \caption{\label{fig:scaltoys} $\abs{ \kappa_{ggh}(m^2_{ZZ})/
        \kappa_{ggh}(m^2_{h})}^2$ as a function of $m_{ZZ}$ for color
      triplet scalar degrees of freedom with $m_s=50$~GeV~(blue) and
      $m_s=350$~GeV~(orange).}
    \end{center}
\end{figure}

\begin{figure}[!t]
  \begin{center}
    \includegraphics[width=0.45\textwidth]{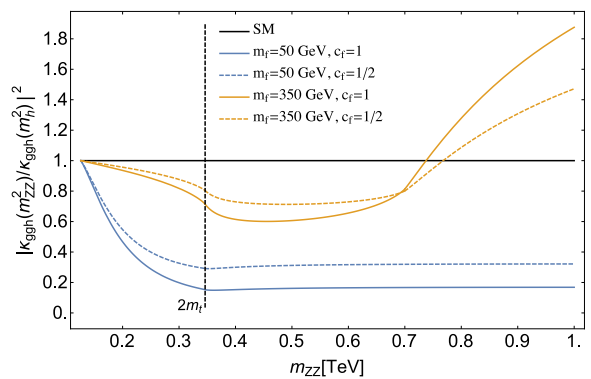}
    \caption{\label{fig:fermtoys} $\abs{ \kappa_{ggh}(m^2_{ZZ})/
        \kappa_{ggh}(m^2_{h})}^2$ as a function of $m_{ZZ}$ for color
      triplet fermionic degrees of freedom with $m_f=50$~GeV~(blue)
      and $m_f=350$~GeV~(orange).}
    \end{center}
\end{figure}

The $ggh$ amplitude relative to the SM one is given
by
\begin{align} 
  \label{eq:kappatoy}
	\kappa_{ggh}(\sh) \simeq
&	\bigg[ 
	\frac{3}{2}\sum_f C(r_f) c_f A_f (\tau_f)
	+\frac{3}{2}\sum_sC(r_s)c_s  A_s(\tau_s)    \nonumber\\
&	+ c_g\frac{3}{\sqrt{2}}\frac{v^2 }{f^2}\frac{y_t^2}{g_\rho^2}\bigg]
	\times \frac{4}{ 3A_t (\tau_t) + 3A_b (\tau_b)} \, ,
\end{align}
where $A_{s,f}$ are the scalar and fermion loop
functions~\cite{Djouadi:2005gi} and $\tau_X =
\sh/(4m^2_X)$. $C(r_X)=1/2$ for the fundamental representation of
$SU(3)$ and the indices $s,f$ run over all scalars and fermions
(i.e. including the SM fermions). We also include an effective $ggh$
interaction as the last term in Eq.~\eqref{eq:kappatoy} that we will
discuss further in Sec.~\ref{sec:eft} below.

\begin{figure}[!t]
  \begin{center}
    \includegraphics[width=0.45\textwidth]{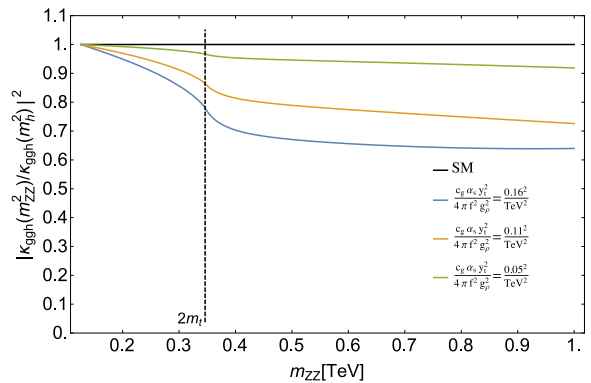}
    \caption{\label{fig:efftoys} $\abs{ \kappa_{ggh}(m^2_{ZZ})/
        \kappa_{ggh}(m^2_{h})}^2$ as a function of $m_{ZZ}$ for the
      operator $H^\dagger H G_{\mu\nu}^a G^{a\mu\nu}$ with varying
      Wilson coefficients~blue,~yellow and~green.}
    \end{center}
\end{figure}

In Figs.~\ref{fig:scaltoys} and \ref{fig:fermtoys} we show the ratio
between the off- and on-shell differential couplings,
$\abs{\kappa_{ggh}(m_{ZZ}^2) /\kappa_{ggh}(m_h^2)}^2 $, as a function
of the $ZZ$ invariant mass. We consider the case of a color-triplet
representation and masses of $m_s,m_f=50,350$~GeV with
$c_s,c_f=1,1/2$. Depending on the size and sign of the BSM couplings,
(a) we can get a cancellation or an enhancement between the SM and the
new physics contributions for the subamplitude that follows from
Fig.~\ref{fig:feyn} (a). If these effects are large we cannot
extrapolate the off-shell region to the on-shell region unless we know
the specifics of the interaction and the particle mass. However, if
the new physics scenario is such that it uniformly converges to the SM
case we can understand the measurement as a probe of the Higgs
width. The dimension six extension of the SM provides an example of
such a scenario as already mentioned in the introduction and shown in
Fig.~\ref{fig:efftoys}. There we show the impact of an effective
operator $H^\dagger H G_{\mu\nu}^a G^{a\mu\nu}$ with a Wilson
coefficient of
\begin{equation}
  \frac{c_g
    g_S^2}{16\pi^2f^2}\frac{y_t^2}{g_\rho^2}=\left( \{0.05,0.11,0.16\}/{\text{
        TeV}}\right)^2\,.
\end{equation}

How realistic is an extension including light degrees of freedom?
In the MSSM, a light scalar can be incorporated as the super partner
of the top. For non-degenerate squark masses, current exclusion limits
for stop searches are depending on several assumptions, e.g. the mass
of the lightest supersymmetric particle~\cite{atlasstops,cmsstops}. Thus, excluding stops with masses in the
100 GeV range categorically is at the moment not possible.

\subsection{Light Degrees of Freedom}
\label{sec:ldof}

\subsubsection*{The MSSM}
As pointed out in the previous section, the MSSM is a candidate model
that can include light scalar degrees of freedom. Furthermore, the
$gg\to ZZ\to 4\ell$ final state will receive additional resonant
contributions from the heavy Higgs partner of the MSSM Higgs
sector. While those contributions are fully included in our
implementation, we will discuss them in detail later in this paper.

To achieve a relatively large mass of $125$ GeV for the lightest
CP-even Higgs boson $h$, while maintaining a light stop, large $A$-terms
are necessary which in turn increase the chiral component of the
stop-Higgs coupling\footnote{Large $A$-terms are constrained by vacuum
  stability requirements~\cite{Reece:2012gi}.}. However, the Higgs
mass constraint can be satisfied by introducing other degrees of
freedom, e.g. as pursued in the NMSSM~\cite{nmssm}, and a large mass
splitting of the two stops can be realized with large soft mass
components $M_{\mathrm{RR},33} \ll M_{\mathrm{RR(LL)},ii}$ or
$M_{\mathrm{LL},33} \ll M_{\mathrm{RR(LL)},ii}$ without inducing a
large Higgs-stop coupling. Therefore, the limits we discuss in
Sec.\ref{sec:qualnonres} can be realized in the (N)MSSM.

\begin{figure}[!t]
  \begin{center}
    \includegraphics[width=0.48\textwidth]{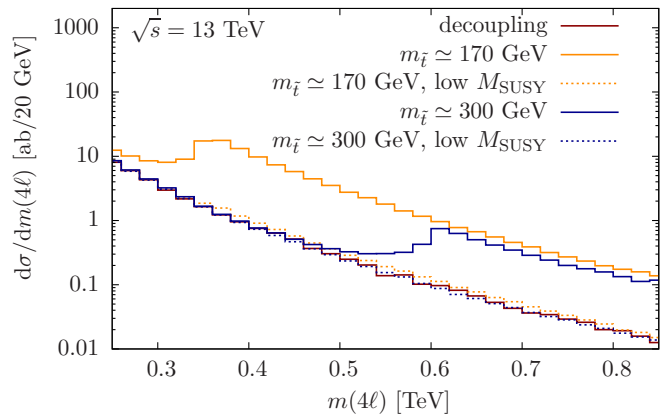}
    \caption{\label{fig:susy} High invariant mass region of $pp\to ZZ
      \to e^+e^-\mu^+\mu^-$ in the (N)MSSM for different choices of
      $M_{\text{SUSY}}$ and stop masses. For details see text.}
    \end{center}
\end{figure}

We do not delve into the details of non-minimal SUSY model-building,
but we want to stress the crucial points that phenomenologically
impact searches at large $m(4\ell)$ from a slightly different angle
compared to the previous section: Since the stop contributions obtain
a chiral component which can be large as a function of the MSSM
parameters $\mu,A_t,$~and $\tan\beta$~\cite{Djouadi:2005gi},
additional thresholds in diagrams of type Fig.~\ref{fig:feyn}~(a) can
impact the high invariant mass tail~\cite{bsm}. We stress that limits
on stops from direct searches highly depend on $m_{\chi^0}$
\cite{atlasstops,cmsstops}, assuming prompt $\tilde{t} \to t~\chi^0$
decays. Thus, probing stops via their contributions to loop-induced
processes can allow to set limits in a less model-dependent way.

Eqs.~(\ref{eq:toy}) expressed in terms of Higgs-quark interactions in
the MSSM yields the coefficients~\cite{Djouadi:2005gi}
\begin{align}
&	c_u=\cos\alpha/\sin\beta \, , \quad
	c_d=-\sin\alpha/\cos\beta \, , 
\end{align}
with $\tan\beta$ being the ratio of the vacuum expectations and
$\alpha$ the neutral scalar mixing angle. For the stop it can be
approximated by
\begin{align}
  \label{eq:stops}
  c_{\tilde{t}} = &\frac{1}{m^2_{\tilde{t}_1} }\Big[
  c_u m^2_t  - \frac{1}{2}s_{2\theta_t}m_t (A_t c_u -  \mu c_d ) \nonumber\\
  &-\frac{1}{6} m^2_Z s_{\alpha+\beta}\left(3-4s^2_W + (-3+8s^2_W)
    s^2_{\theta_t} \right)\Big] \, ,
\end{align}
where $s_X\equiv\sin(X)$, $c_X\equiv\cos(X)$ and $\sin(2\theta_t)=
2m_t(A_t-\mu \cot\beta)/(m^2_{\tilde{t}_1} - m^2_{\tilde{t}_2})$ is
the stop mixing angle with the trilinear coupling $A_t$.

To understand the quantitative effects, we choose $\mu=100~\text{GeV}$
throughout and consider
\begin{alignat}{5}
  \hbox{(i)}\, &M_{\text{SUSY}}=1.0~\text{TeV},~\tan\beta=2 \,,\\
  \hbox{(ii)}\, &M_{\text{SUSY}}=0.5~\text{TeV},~\tan\beta=2 \,.
\end{alignat}
We assume degenerate soft-mass terms $M_{\mathrm{RR,LL}} =
M_{\text{SUSY}}$ and vary $A_t$ such to obtain $m_{\tilde t}\simeq
170~\text{GeV}$ and $m_{\tilde t}\simeq 300~\text{GeV}$. Hence, larger
$M_\text{SUSY}$ results in larger $A_t$ and therefore larger
Higgs-stop couplings, see Eq.~(\ref{eq:stops}).  The high invariant
mass region in $pp\to ZZ\to 4\ell$ can become an efficient indirect
probe of the existence of light stops provided a non-negligible
Higgs-stop coupling. The latter is phenomenologically preferred to
achieve the relatively large $m_h \simeq 125$ GeV.

We show the different $m_{ZZ}$ distributions for those parameter
choices in Fig.~\ref{fig:susy}, keeping $m_h=125~\text{GeV}$
fixed. Constraints on low stop masses in this particular parameter
range of the (N)MSSM can be formulated in the absence of a
stop-induced threshold for $m_{ZZ} > m_h$. As demonstrated in
Fig.~\ref{fig:susy}, the effects quickly decouple with larger stop
masses and smaller values of $A_t\lesssim 1~\text{TeV}$.

\subsection{Effective Field Theory}
\label{sec:eft}
 
Higgs effective field theory has gained a lot of attention in the past
and recently~\cite{dim6,dim6e,dim6r,dim6g,dim6gg1,dim6gg1,dim6f} and there is a
rich phenomenology of anomalous Higgs couplings in $gg\to ZZ \to
e^+e^- \mu^+\mu^-$ production. To keep our discussion as transparent
as possible we will choose the convention of~\cite{dim6g} in the
following:
\begin{widetext}
  \vspace{-0.5cm}
  \begin{align} 
    {\cal L}_{\text{SILH}}  = 
    &\frac{c_H}{2f^2}\partial^\mu \left( H^\dagger H \right) \partial_\mu
    \left( H^\dagger H \right)
    + \frac{c_T}{2f^2}\left (H^\dagger {\overleftrightarrow { D^\mu}} H
    \right)  
    \left(   H^\dagger{\overleftrightarrow D}_\mu H\right) 
    - \frac{c_6\lambda}{f^2}\left( H^\dagger H \right)^3  
    + \left( \frac{c_yy_f}{f^2}H^\dagger H  {\bar f}_L Hf_R +{\rm
        h.c.}\right)  \nonumber\\ 
    &+\frac{ic_Wg}{2m_\rho^2}\left( H^\dagger  \sigma^i \overleftrightarrow
      {D^\mu} H \right )( D^\nu  W_{\mu \nu})^i 
    +\frac{ic_Bg'}{2m_\rho^2}\left( H^\dagger  \overleftrightarrow {D^\mu}
      H \right )( \partial^\nu  B_{\mu \nu})  
    +\frac{ic_{HW} g}{16\pi^2f^2}
    (D^\mu H)^\dagger \sigma^i(D^\nu H)W_{\mu \nu}^i  \nonumber\\ 
    &+\frac{ic_{HB}g^\prime}{16\pi^2f^2}
    (D^\mu H)^\dagger (D^\nu H)B_{\mu \nu}
    +\frac{c_\gamma {g'}^2}{16\pi^2f^2}\frac{g^2}{g_\rho^2}H^\dagger H
    B_{\mu\nu}B^{\mu\nu} 
    +\frac{c_g g_S^2}{16\pi^2f^2}\frac{y_t^2}{g_\rho^2}H^\dagger H
    G_{\mu\nu}^a G^{a\mu\nu}\,,
    \label{lsilh}
  \end{align}
\end{widetext}
with $H^\dagger \overleftrightarrow { D^\mu} H= H^\dagger D^\mu H -
(D^\mu H^\dagger)H$. It is worth pointing out that the operator basis
is completely identical to a general dimension six extension of the SM
Higgs sector~\cite{dim6r}, and differs from it by a bias on the Wilson
coefficients that can be motivated from an approximate shift symmetry
related to the interpretation of the Higgs as pseudo-Nambu Goldstone
boson~\cite{dim6g}. This bias suppresses certain operators relative to
others, and the differential cross section will mostly depend on a
subset of Wilson coefficients for identically chosen coefficients
$c_i$ in Eq.~\eqref{lsilh}. In a particular BSM scenario this can or
might not be true; we simply adopt the language of~\cite{dim6g} to
illustrate the quantitative impact of a highlighted set of dimension
six operators, while our numerical implementation incorporates all
operator structures of Eq.~\eqref{lsilh}. We work with a canonically
normalized and diagonalized particle spectrum that, after appropriate
finite field and coupling renormalization, does not modify the $gg\to
ZZ$ continuum contribution (this has been checked numerically and
analytically).

\begin{figure}[!t]
  \begin{center}
    \subfigure[]{\includegraphics[width=0.45\textwidth]{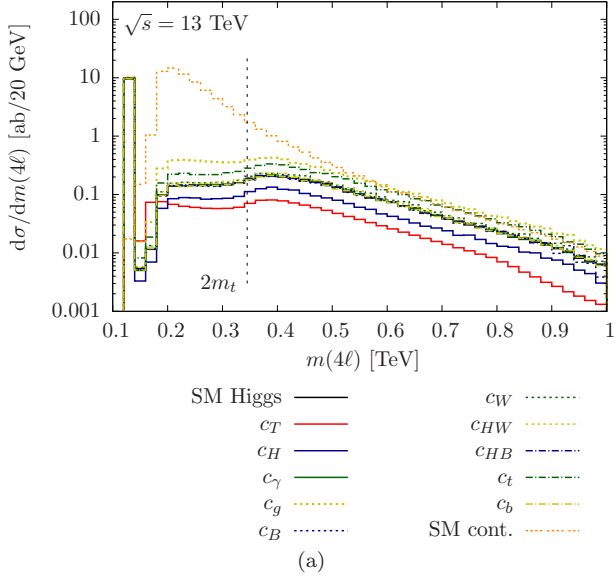}}
    \subfigure[]{\includegraphics[width=0.45\textwidth]{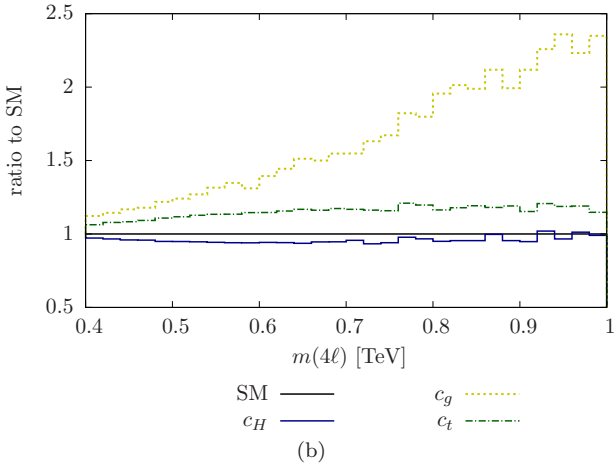}}
    \caption{\label{fig:eft} (a) Individual cross section
      contributions to $p(g)p(g)\to ZZ \to e^+e^-\mu^+\mu^-$ as a
      function of the parameters of Eq.~\eqref{lsilh}, subject to the
      constraint $\mu^{\text{on}}_{ZZ}=1$. Note that $c_T$ shifts
      $m_Z$ away from its SM value, which is tightly constrained by
      the $T$ parameter~\cite{Peskin:1991sw}. The modification of the
      intermediate $Z$ boson mass is not reflected in the SM continuum
      distribution, which is purely SM. We also show the impact of the
      dominant ${\cal{L}}_{\text{SILH}}$ operators in the full cross
      section, taking into account all interference effects, relative
      to the SM expectation in panel (b). We choose Wilson
      coefficients of size $c_i v^2/f^2 \simeq 0.25$ in both panels.}
    \label{fig:wqeft}
  \end{center}
\end{figure}

We do not consider dipole operators of the form $\sim \bar q
\sigma^{\mu\nu} \sigma^i H^c q\, W^i_{\mu\nu}$ which will impact the
continuum production of $gg \to ZZ \to 4\ell$ and $\bar q q \to
ZZ$. New physics contributions to the latter processes need to be
treated independently in a concrete experimental analysis and is
beyond the scope of our work. For demonstration purposes we choose
\begin{equation}
  f =m_\rho=5~\text{TeV},~g_\rho=1\,.
\end{equation}
and $c_i v^2/f^2 \simeq 0.25$ for the $m_{ZZ}$ spectra of
Fig.~\ref{fig:wqeft}.

From Fig.~\ref{fig:wqeft}, it becomes apparent that the high invariant
mass region has an excellent sensitivity to the dimension six
operators of Eq.~\eqref{lsilh}. We have chosen a SM signal strength
$\mu^{\text{on}}_{ZZ}=1$ which selects a region in the space of Wilson
coefficients~\cite{dim6gg1}. This region can be further constrained by
including complementary information from a measurement of
$m_{ZZ}\gtrsim 330~\text{GeV}$ region~\cite{fermilab1,bsmh,bsmh2}. This
allows us to formulate the Higgs width as a function of the relevant
dimension six operator coefficients through correlating
Eqs.~\eqref{eq:1a} and \eqref{eq:1b}. Note that operator
mixing~\cite{trott1,espinosa} is anticipated to impact the
phenomenology of this Lagrangian at the 10\% level if scales are
vastly separated~\cite{trott2,Englert:2014cva}. Hence, the comparison
of on- and off-shell measurements is direct
$c_i(m_h)=c_i(m_{ZZ}>330~\text{GeV})$. If we invoke the operator
coefficient bias and of Eq.~\eqref{lsilh} focus on a tree-level $T$
parameter $T=0$, the dominant operator coefficients that are probed in
the off-shell region are $c_H,c_g,c_t$.

A targeted analysis of how far these parameters can be constrained at
the LHC has been presented in Ref.~\cite{bsmh2}; a question that
remains worth addressing in this context, however, is the impact of
the off-shell measurement in comparison to Higgs measurements in other
channels such as associated Higgs \cite{Butterworth:2008iy,asso} and
Higgs+jet \cite{monojet1} production.
  
\begin{figure}[!t]
  \begin{center}
    \includegraphics[width=0.45\textwidth]{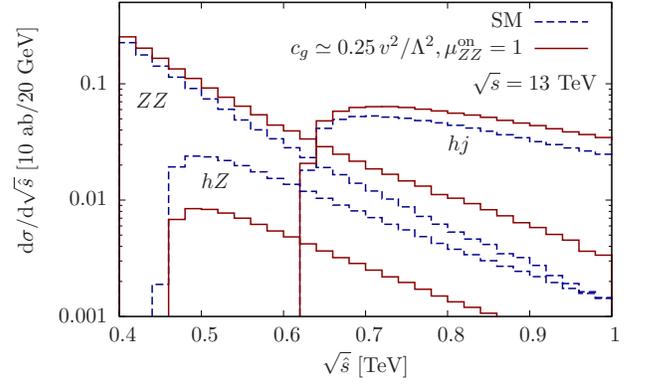}
    \caption{\label{fig:eftcomp} Comparison of the off-shell
      measurement of $pp\to ZZ\to$ light leptons with associated
      $pp\to hZ\to b\bar b \ell^+\ell^-$ ($\ell=e,\mu$) and $pp\to hj
      \to \tau^+\tau^-$.}
  \end{center}
\end{figure}

In the following we input the SM-like signal strengths in the $pp\to
ZZ$ channel since direct measurements in the latter channels are not
available at 8 TeV. The signal distributions for a representative
operator choice $c_g\simeq 0.25 v^2/\Lambda^2$ is given in
Fig.~\ref{fig:eftcomp} and we use eHdecay~\cite{dim6gg1} to compute
the modified branching ratios, inputing the the bigger Higgs width to
achieve $\mu^{\rm on}_{ZZ}=1$. The different thresholds and
normalizations in Fig.~\ref{fig:eftcomp} reflect the signal regions
and selection efficiencies as documented in the
literature~\cite{asso,monojet} due to $b$-tagging, $\tau$
reconstruction and subjet techniques.

It should be noted that associated Higgs and Higgs+jet production are
plagued with large backgrounds as opposed to the experimentally clean
$ZZ\to 4\ell$ signature\footnote{For instance, a measurement of the
  off-shell cross section is already available with the 8 TeV data set
  although the inclusive signal cross section is significantly smaller
  compared to $Z$-associated and jet-associated Higgs production}, the
signal-to-background ratio in e.g. $pp\to hj\to \tau^+\tau^-$ is of
the order of 0.1 \cite{monojet}. A measurement of the differential
distributions as shown in Fig.~\ref{fig:eftcomp} in these channels
will be complicated: While the acceptance in the fully leptonic $ZZ$
final state at large invariant four-lepton masses is close to unity
\cite{atlaswidth,cmswidth}, the signal rates in associated and monojet
production are vastly reduced (for details see e.g. \cite{asso} and
\cite{monojet}). Therefore, off-shell measurements in the $pp\to ZZ$
channel will not only provide crucial information to limit the
presence of higher dimensional operators but also provide
complementary information, in particular due to a larger kinematically
accessible phase space range.

\begin{figure*}[!t]
  \begin{center}
    \subfigure[]{\includegraphics[width=0.45\textwidth]{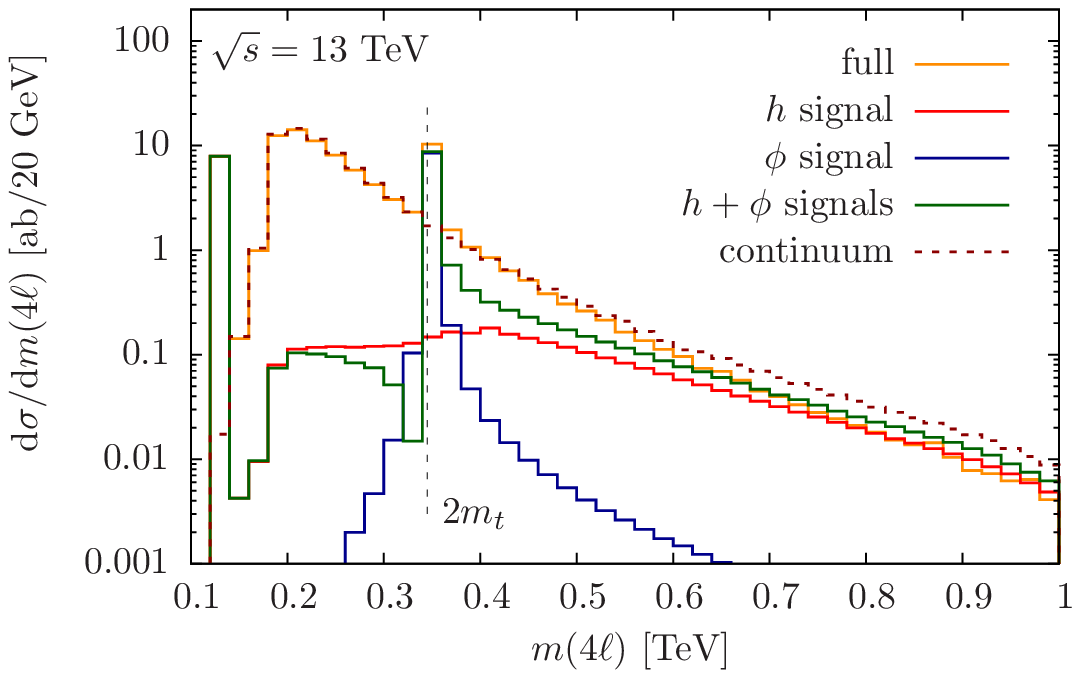}}
    \hfill
    \subfigure[]{\includegraphics[width=0.45\textwidth]{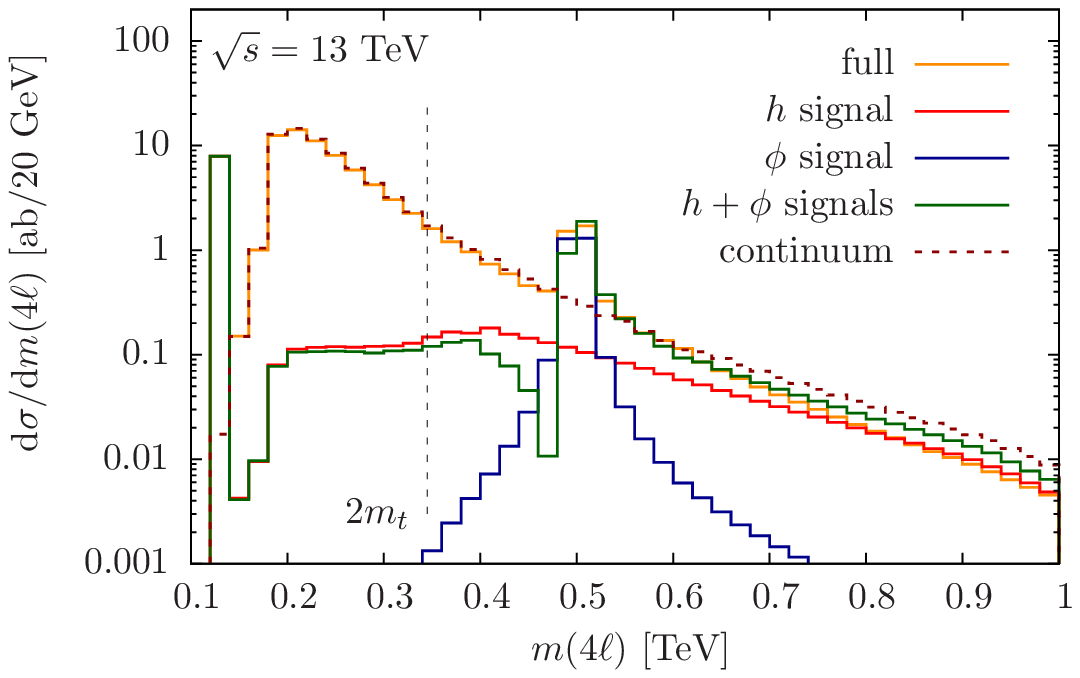}}
    \caption{ \label{fig:wq} Individual and combined ``signal''
      contributions, as well as full differential cross sections in
      the portal-extended SM for $\cos^2\chi=0.9$ and two choices of
      heavy boson masses $m_\phi=350~\text{GeV}$ and
      $m_\phi=500~\text{GeV}$ for SM-like width values
      $\Gamma_\phi(m_\phi)= 0.1\, \Gamma_h^{\text{SM}}(m_\phi)$.}
  \end{center}
\end{figure*}

\section{Resonant BSM Physics}
\label{sec:res}
In contrast to the non-resonant physics scenarios discussed in the
previous sections, we can imagine the off-shell measurement to be
impacted by the presence of additional iso-singlet scalar
resonances. To work in a consistent framework, we will focus on
so-called Higgs portal scenarios~\cite{portal_orig} in the following,
which directly link the presence of new scalar states to a universal
Higgs coupling suppression. We focus on the minimal extension of the
Higgs sector
\begin{equation} 
  \label{eq:portal}
  {\cal{L}}_{\text{Higgs}}=\mu^2 |H|^2 - \lambda |H|^4 +\eta |H|^2
  |\phi|^2  +\tilde\mu^2 |\phi|^2 - \tilde \lambda |\phi|^4\,.
\end{equation}
If both the Higgs doublet $H$ and the extra singlet $\phi$ obtain a
vacuum expectation value, the $\eta$-induced linear mixing introduces
a characteristic mixing angle $\cos\chi$ to single Higgs phenomenology
via rotating the Lagrangian eigenstates (${\cal{L}}$) to the mass
eigenbasis (${\cal{M}}$)\footnote{Multi-Higgs phenomenology can be
  vastly different~\cite{choiportal}.}
\begin{equation}
  \left(\begin{matrix} h \\ \phi \end{matrix} \right)_{\cal{L}} = 
  \left(\begin{matrix} \cos \chi & -\sin\chi \\ \sin\chi & \cos\chi  \end{matrix} \right)
  \left(\begin{matrix} h \\ \phi \end{matrix} \right)_{\cal{M}}\,.
\end{equation}
Consequently, we have two mass states with a SM-like phenomenology;
such models have been studied in detail and we refer the reader to the
literature~\cite{portal0,portal,choiportal}.

\begin{figure}[!t]
  \begin{center}
    \subfigure[]{\includegraphics[width=0.45\textwidth]{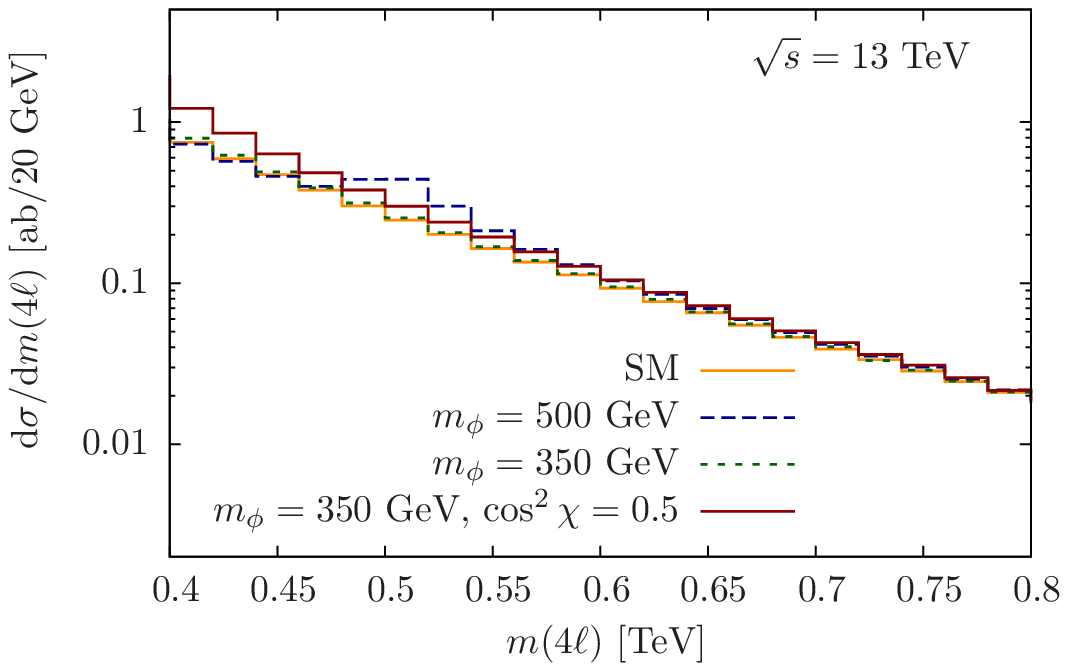}}\\
    \subfigure[]{\includegraphics[width=0.45\textwidth]{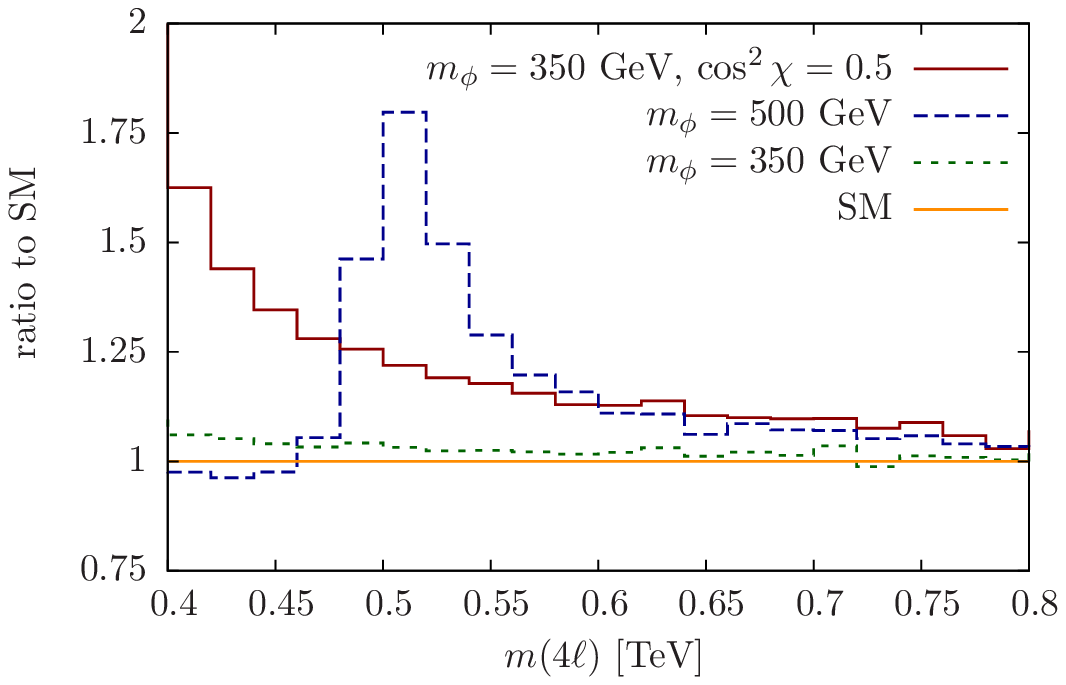}}
    \caption{\label{fig:wq2} Full differential cross section at high
      invariant masses for the SM and the two choices of $m_\phi$. For
      $m_\phi=500~\text{GeV}$ we choose $\Gamma_\phi=40~\text{GeV}$ to
      enhance visibility for the ratio plot shown in the lower panel.}
  \end{center}
\end{figure}

We focus on scenarios
\begin{alignat}{5}
  m_h=125~\text{GeV} &:\quad \text{coupling suppression $\cos\chi$} \\
  m_\phi>m_h &:\quad \text{coupling suppression $\sin\chi$}
\end{alignat}
and keep the Higgs width identical to the SM (this could be
facilitated by another portal interaction to light SM-singlet
states). This will modify the on-shell Higgs phenomenology and we
choose $\mu^{\text{on}}_{ZZ}=\cos^4\chi=0.81$, which is within the
$H\to ZZ$ limits as reported in latest coupling fits in the $ZZ$
category~(see e.g.~\cite{hzz}). This choice is also consistent with
the non-observation of a heavy Higgs-like particle with a signal
strength of $\sim 10\%$ of the SM expectation in a region where the
narrow width approximation is valid (see e.g. recent searches by
CMS~\cite{Chatrchyan:2013yoa}) and limits set by electroweak precision
constraints; see also \cite{bsmres2} for a detailed discussion of
currently allowed parameter range, and \cite{portal0} on constraints
that can be obtained by measuring the heavy Higgs boson

Since the light Higgs width quickly decouples this choice is
irrelevant for the phenomenology at high invariant mass. To keep our
discussion transparent, we choose a trivial hidden sector
phenomenology by using
\begin{equation}
  \Gamma_\phi(m_\phi)= \sin^2\chi \, \Gamma_h^{\text{SM}}(m_\phi)
\end{equation}
in the following. The results for two representative choices of
$m_\phi$ are shown in Fig.~\ref{fig:wq}.

The structure in the ``$H+\phi$'' signal results from a destructive
interference of the Higgs diagrams in the intermediate region $m_h <
\sqrt{\sh} \lesssim m_\phi$ as a consequence of the propagator
structure and will depend on how we formulate the Higgs width
theoretically~\cite{Hwidth}.\footnote{A survey of dip structures in
  cross sections has been presented in
  Refs.~\cite{Bai:2014fkl,Stuart:1991xk}.} From a phenomenological
perspective this structure is numerically irrelevant.

Apart from the obvious additional resonance, we do not find a notable
deviation from the SM away from the Breit-Wigner ``turn on'' region
$m(4\ell)\gtrsim m_\phi$. Away from all $s$-channel particle
thresholds, i.e. for invariant masses $m(4\ell)\gg m_\phi$, the
amplitude becomes highly resemblant to the SM amplitude as a
consequence of the linear mixing: If we write the SM top-triangle
subamplitude as ${\cal{C}}(\hat s,m^2_t)$ and remove the $Z$ boson
polarization vectors, we have an amplitude
\begin{multline}
  \label{eq:cos}
  {\cal M}^{\mu\nu} = g^{\mu\nu} {\cal{C}}(\hat s,m_t^2) \\
  \times\left( {\cos^2\chi\over \sh - m_h^2 +i m_h\Gamma_h} +
    {\sin^2\chi\over \sh - m_\phi^2 +i m_\phi\Gamma_\phi} \right)\\ \to
  {g^{\mu\nu}\over \sh}{\cal{C}} (\hat s,m_t^2)\quad\hbox{for $\sh\gg
    m_h^2,m_\phi^2$,}
\end{multline}
which is just the SM contribution evaluated at large $\sqrt{\hat s}$. This
qualitative argument is numerically validated for the full cross
section in Fig.~\ref{fig:wq2}. The differential $m_{ZZ}$ distribution
approaches the SM distribution rather quickly, especially because
consistency with the 125 GeV signal strength measurements and
electroweak precision data~\cite{elwp} imposes a
hierarchy $\cos^2\chi \gg \sin^2\chi$.

Eq.~\eqref{eq:cos} suggests that the more interesting parameter choice
for modified interference effects at large invariant masses is a
larger mixing. In this case, however, the Higgs on-shell phenomenology
would vastly modified too. Larger values of $\sin^2\chi$ also imply
tension with electroweak precision data and direct search constraints,
unless we give up the simplified model of Eq.~\eqref{eq:portal}. This
is beyond the scope of this work. Quantitatively a larger mixing only
shows a moderate increase for $m(4\ell)\gtrsim 400$~GeV (we include a
maximum mixing angle $\cos^2\chi=0.5, m_\phi=350~\text{GeV}$ to
Fig.~\ref{fig:wq2}), which results from Breit-Wigner distribution of
the state $\phi$; for maximal mixing this has a larger signal strength
compared to the $\cos^2\chi=0.9$ scenario.

In summary, we conclude that the basic arguments that have been used
in the interpretation of SM
measurements~\cite{Kauer:2013qba,ciaran,mit,Dixon:2013haa2,cmswidth,atlaswidth}
remain valid in this minimal resonant extension of the SM Higgs
sector. Our analysis straightforwardly generalizes to the two Higgs
doublet model~\cite{2hdm} and the $n$HDM~\cite{nhdm}.

\section{Summary and Conclusions}
Measurements at large momentum transfers as a probe of non-decoupling
off-shell Higgs contributions provide an excellent testing ground of
various scenarios of BSM physics.

In this paper we have further examined the validity of the
interpretation of off-shell measurements as a probe of the Higgs total
width. In combination with a signal strength
$\mu^{\text{on}}_{ZZ}\simeq 1$, we motivate the double ratio
$R(m_{ZZ})$ of Eq.~\eqref{eq:doubler} as guideline for when this
interpretation is valid, namely $R\simeq 1$ within uncertainties.

Furthermore, measurements at large invariant $ZZ$ masses in $pp\to
ZZ\to 4\ell$ at the LHC run~2 will have significant impact on searches
for BSM physics far beyond the interpretation in terms of the Higgs'
width. We have discussed a wide range of BSM scenarios as examples
that highlight this fact. In particular, we have provided a
quantitative analysis of the high invariant mass region of $pp\to
ZZ\to 4\ell$ in the context of the MSSM, a general dimension six
extension of the SM Higgs sector, and resonant phenomena within Higgs
portal scenarios.

Generic to all BSM scenarios is the model-dependence of the off-shell
region. If we observe an excess in the future in the high $m_{ZZ}$
region, the interpretation of such an observation is not necessarily
related to the Higgs but could be a general effect of the presence of
new TeV-scale dynamics. In particular, the ``off-shell signal
strength'' has no relation to on-shell Higgs properties such as the
width or even Higgs couplings, unless imposed by a choice of a
particular class of BSM scenarios such as Eq.~\eqref{lsilh}. An
example of that, which we have not discussed in further detail are
electroweak magnetic operators or an additional broad and heavy $Z'$
boson, that can impact the $q\bar q$-induced production channels in a
way that is a priori unrelated to the Higgs sector.

\acknowledgments We thank Ian Low for suggesting a quantitative
analysis of the interference effects in the portal-extended SM and
Gilad Perez and Andreas Weiler for valuable discussions.
CE is supported by the Institute for Particle Physics Phenomenology
Associateship program.
MS thanks the Aspen Center for Physics for hospitality while part of
this work was completed. This work was supported in part by the
National Science Foundation under Grant No. PHYS-1066293.


\end{document}